\documentclass[12pt]{iopart}

\usepackage{iopams}  
\usepackage{graphicx}
\usepackage{bm}
\usepackage{hyperref}
\usepackage{xcolor}
\begin{document}

\title[]{Oxygen vacancy in ZnO-$w$ phase: pseudohybrid Hubbard density functional study}

\author{Ivan I. Vrubel$^1$, Anastasiia A. Pervishko$^{1,2}$, Dmitry Yudin$^{1,2}$, Biplab Sanyal$^2$, Olle Eriksson$^{2,3}$ and Piotr A. Rodnyi$^4$}
\address{$^1$ Skolkovo Institute of Science and Technology, Moscow 121205, Russia}
\address{$^2$ Department of Physics and Astronomy, Uppsala University, Box 516, SE-751 20 Uppsala, Sweden}
\address{$^3$ School of Science and Technology, \"Orebro University, SE-701 82 \"Orebro, Sweden}
\address{$^4$ Peter the Great St. Petersburg Polytechnic University, St. Petersburg 195251, Russia}
\ead{ivanvrubel@ya.ru}

\vspace{10pt}
\begin{indented}
\item[]March 2020
\end{indented}

\begin{abstract}
The study of zinc oxide, within the homogeneous electron gas approximation, results in overhybridization of zinc $3d$ shell with oxygen $2p$ shell, a problem shown for most transition metal chalcogenides. This problem can be partially overcome by using LDA+$U$ (or, GGA+$U$) methodology. However, in contrast to the zinc $3d$ orbital, Hubbard type correction is typically excluded for the oxygen $2p$ orbital. In this work, we provide results of electronic structure calculations of an oxygen vacancy in ZnO supercell from {\it ab initio} perspective, with two Hubbard type corrections, $U_{\mathrm{Zn}-3d}$ and $U_{\mathrm{O}-2p}$. The results of our numerical simulations clearly reveal that the account of $U_{\mathrm{O}-2p}$ has a significant impact on the properties of bulk ZnO, in particular the relaxed lattice constants, effective mass of charge carriers as well as the bandgap. For a set of validated values of $U_{\mathrm{Zn}-3d}$ and $U_{\mathrm{O}-2p}$ we demonstrate the appearance of a localized state associated with the oxygen vacancy positioned in the bandgap of the ZnO supercell. Our numerical findings suggest that the defect state is characterized by the highest overlap with the conduction band states as obtained in the calculations with no Hubbard-type correction included. We argue that the electronic density of the defect state is primarily determined by Zn atoms closest to the vacancy.
\end{abstract}

\vspace{2pc}

\submitto{\JPCM}

\section{Introduction}
Thanks to its unique electronic, magnetic and optical properties and in the response to the actual industrial challenges, zinc oxide (ZnO) has remained in the focus of intensive theoretical and experimental studies for several decades \cite{Kohan00,Klingshirn2007,Janotti2009,Radzimska2014}. A delicate interplay between electrical and mechanical properties endows its crystalline structure, that lacks inversion symmetry, with highly pronounced piezoelectric properties \cite{Hill2000}. The wurzite phase of ZnO, is a typical ionic semiconductor with the bandgap ranging from 3.44~eV at 4~K to 3.37~eV at 300~K. This compound is known to host excitons with the binding energy of 60~meV even at room temperature \cite{ellmer08,ozgur05}, prompting its utility as a source for UV lasing \cite{Bagnall1997} with the variety of applications in sensor development, along with luminophore and scintillation techniques \cite{Rodnyi2011,Gorokhova2008}. Moreover, long spin coherence times \cite{Ghosh2005} and minor resistivity of nonstoichiometric undoped thin films of ZnO \cite{Look1999,Zhang2001} favor the use of this material for faster data processing. Besides excitonic luminescence (3.35~eV, $\tau$=0.7~ns) in the vicinity of the absorption edge, samples of ZnO are known to reveal the so-called green luminescence, that is characterized by broadband emission spectrum with the characteristic decay time of the order of $\mu$s. Noteworthy, the green luminescence is usually associated with point defects \cite{Khranovskyy12}, mostly in the form of oxygen vacancies \cite{angelis11,azpiroz11}. Practically, annealing in oxygen-rich atmosphere or implanting In and Ga oxides quenches the green luminescence \cite{chernenko18}, whereas it is robust when introducing metallic In and Ga instead \cite{bourret09}.

From a theoretical viewpoint, the band structure of zinc oxide was first approached using the Green's function KKR method \cite{Rossler1969}, and subsequently examined by pseudopotential calculations where zinc $3d$ electrons were placed in the core \cite{Bloom1973,Chelikowsky1977}. As it was shown, first principles calculations are computationally demanding since both zinc and oxygen appear to be not well suited for the construction of a pseudopotential \cite{Schroer1993}, and, furthermore, the routinely used local density (LDA) and generalized gradient approximations (GGA) give rise to overhybridized zinc $3d$ and oxygen $2p$ orbitals. Additionally, defect states in zinc oxide, e.g., the ones formed by oxygen vacancies, cannot be reproduced satisfactorily in the form of localized states in a supercell. To subdue these limitations, more involved techniques \cite{morales17} have been developed, such as methods that take into account Hubbard correction \cite{goh17,janotti05,Erhart06,Freysoldt14,Boonchun11,Janotti11,Oba10} (within DFT+$U$ methodology \cite{liechtenstein95}) and hybrid functionals \cite{betzinger10,bashyal18,muscat01,adamo99,Wrubel09,king09}, as well as Green's functions based approach \cite{zhang16,Sarsari13}. In this work, we provide the results of electronic structure calculations for a ZnO supercell with and without oxygen vacancy. We show that applying Hubbard correction to both Zn $3d$ and O $2p$ orbitals within DFT+$U$ method allows to reproduce correctly the band structure in a quantitative way, which enables a detailed study of the localized electronic density of an oxygen vacancy.

\section{Computational methods}

\subsection{Methodology}

The unit cell of ionic ZnO in a wurtzite phase is a hexagonal close-packed lattice with each anion being surrounded by four cations positioned at tetragonal sites and vice versa, with experimentally determined lattice constants $a=3.250$~\AA\, and $c=5.207$~\AA\, \cite{kisi89}. The occupied polyhedra are not ideal, which is accounted by the internal ordering parameter $u=0.3817$, that defines the length of the anion-cation bond parallel to the $c$ axis in the equilibrium geometry \cite{kisi89}. In this work, to carry out first principles calculations we employ Quantum ESSPRESSO package \cite{qe09,qe17} based on plane-wave density functional theory (DFT). We proceed to a further analysis of the resulting Kohn-Sham wave functions by evaluating projected density of states (PDOS). For postprocessing investigations, we use \texttt{XCrySDen} visualization package with the cutoff energy for the basis set at 90~Ry. The exchange correlation potential is approximated by GGA as implemented by Perdew, Burke and Ernzerhof (PBE) \cite{pbe96}. Noteworthy, the standard Quantum~ESPRESSO  pseudopotentials Zn.pw-mt\_fhi.UPF and O.pw-mt\_fhi.UPF of core shells with explicit 4$s^2$3$d^{10}$ (zinc) and 2$s^2$2$p^4$ (oxygen) electronic structure are employed. The single particle potential is corrected with the on-site Hubbard terms (in the spirit of DFT+$U$). Unless otherwise stated, these values are fixed for both atoms $U_{\mathrm{Zn}-3d}=U_{\mathrm{O}-2p}=8$~eV. For calculations of ZnO unit cell we adopt the $6\times6\times6$ grid for sampling the Brillouin zone \cite{vrubel19}. We address the effect of a single O vacancy, and its influence on the electronic structure, by considering a ZnO supercell with 107 atoms (3$\times$3$\times$3 unit cells), using automatic 2$\times$2$\times$2 Monkhorst-Pack grid \cite{monkhorst76} centered at the $\Gamma$-point of the Brillouin zone.

\subsection{Effect of Hubbard correction}

Standard and modified DFT-based techniques have been extensively used for a few decades to address the properties of ZnO \cite{azpiroz11,betzinger10,bashyal18,hu08,janotti06,lim12,oba08,paudel08,agapito15,topsakal09,flores18,clark10,gori10,oba11,lany10}. Besides a too large hybridization between zinc and oxygen orbitals, calculations based on the DFT methodology suffer from yielding an underestimated bandgap (0.8~eV as compared to experimental value of 3.37~eV). This shortcoming originates from the fact that the homogeneous electron gas approximation is not suitable to correctly capture  the electronic structure of the zinc semicore $d$ shell. Strictly speaking eigenvalues of DFT master equation, the Kohn-Sham equation, in the standard approximations (LDA or GGA) do not correspond to quasiparticle excitations. For this reason the bandgap of LDA or GGA approaches is often underestimated when compared to experimental values. Significant improvements can be achieved by using hybrid functionals (e.g., PBE0), DFT+$U$, or even $GW$ methods. Motivated by the recent studies in \cite{agapito15}, in this work we adopt a pseudohybrid Hubbard DFT approach, where Hubbard corrections, $U_{\mathrm{Zn}-3d}$ and $U_{\mathrm{O}-2p}$, have to be applied for both zinc and oxygen atoms and we explore the role of oxygen vacancies on the electronic structure of zinc oxide, with particular focus on the effect of the $U_{\mathrm{Zn}-3d}$ parameter that allows one to separate Zn $3d$ and O $2p$ states, whereas $U_{\mathrm{O}-2p}$ leads to the gap opening up to larger values.

Since the proposed computational method requires a set of auxiliary parameters, the choice of their values has to be discussed in detail. Initially, DFT$+U$ methodology was introduced to correct calculations of the systems with localized orbitals, where one-site electron-electron interaction is poorly described by conventional parametrisations of exchange-correlation functional. Starting with Hubbard correction applied only to Zn $3d$ shell, the reported values vary significantly from one approach to the other; e.g., in Refs.~\cite{janotti06,janotti07} the utilized $U_{\mathrm{Zn}-3d}$ parameter is assessed through the calculations of the dielectric constant using linear response theory. Using this approach yields the Hubbard correction to be equal to 4.7~eV. This value results in the small change of the ZnO bandgap from 0.8~eV (within standard LDA calculation) to 1.5~eV. Similar behavior was observed in Ref.~\cite{lany08}, where the Hubbard correction was set to be 7~eV for the $3d$ orbital. In Ref.~\cite{lany08}, the alternative idea of using mixed $s/d$ correction to Zn was also discussed. The proposed values for $U_{\mathrm{Zn}-d}$=4~eV and $U_{\mathrm{Zn}-s}$=38~eV provide a good fitting of the ZnO band structure. However, the later value looks excessive and can be attributed to the negligible occupation of Zn $s$ orbital compensated by the introduced parameter that subsequently affects the material band structure. The similar picture was reported in Ref.~\cite{paudel08} where the $U_{\mathrm{Zn}-s}$ parameter equals 44~eV.
The pure theoretical approach of Ref.~\cite{agapito15}, that also triggered the present study, reports optimal values of $U_{\mathrm{Zn}-3d}$ and  $U_{\mathrm{O}-2p}$ to be equal to 12.8~eV and 5.29~eV, respectively. It is also worth noting that papers considering either $3d$ or $3d/2p$ corrections often depict the effect of a series of parameters, and their influence on the specific material properties. This is sometimes done without discussing the optimal values during this empirical search (see e.g. \cite{bashyal18,goh17,huang12,ma13}). Meanwhile, in a search of the optimal values of the Hubbard correction for Zn-based system interface, in Ref.~\cite{flores18} a special attention was paid to the obtained band structure and band gap values and similar optimal correction values were reported. 

To summarize, in our study we set both $U_{\mathrm{Zn}-3d}$=$U_{\mathrm{O}-2p}$ to be 8~eV. This is based on an analysis of the calculated material properties, and their dependence on $U$. In order to find optimal values we varied $U_{\mathrm{O}-2p}$ parameter while controlling lattice constants and simultaneously assessing the bandgap to agree with experimental results. Additionally, the $U_{\mathrm{Zn}-3d}$ value was tuned for positioning of Zn $3d$ states to lie approximately 8~eV below the maximum of valence band that agrees with XPS \cite{ley1974} and ARPES \cite{lim12} data. It should also be noted that the finest tuning of the parameters can be obscured by the numerical features of the algorithm performing calculations of the on-site interaction and utilized basis set \cite{agapito15}. As it will be shown below, our results are stable against variations of Hubbard correction, $U$.

\section{Results and discussion}

\subsection{Choice of optimal Hubbard parameters}

\begin{figure}[ht]
\centering
\includegraphics[scale=0.9]{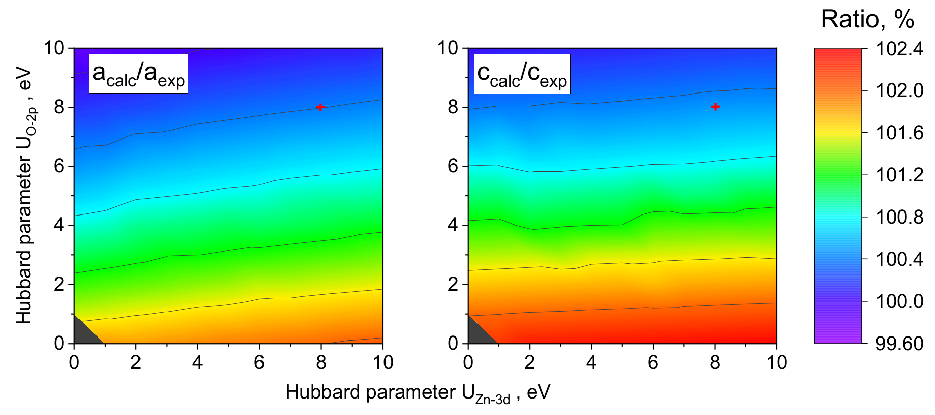}
\caption{Effect of the Hubbard corrections on the values of relaxed lattice parameters, $a$ (left panel) and $c$ (right panel), of the ZnO unit cell with calculated equilibrium geometry. These values are calculated in the form of the phase diagram with $U_{\mathrm{Zn}-3d}$ and $U_{\mathrm{O}-2p}$ varying along horizontal and vertical axes respectively and are further normalized to experimental data \cite{kisi89}. The red cross points at the values used throughout our analysis, $U_{\mathrm{O}-2p}=U_{\mathrm{Zn}-3d}=8$ eV.}
\label{fig1new}
\end{figure}

The calculated equilibrium geometry of a ZnO unit cell, and its dependence on the Hubbard parameters, is shown in figure~\ref{fig1new} in the form of ratio of calculated and experimental values of lattice constants \cite{kisi89}. The calculations clearly demonstrate that variation of the Hubbard correction applied only to Zn $3d$ shell leads to a minor change of equilibrium geometry of the unit cell. In contrast, tuning the Hubbard parameter of O $2p$ electrons results in values of the lattice constant that approach the experimental data. The GGA calculations with a negligible value of $U_{\mathrm{O}-2p}$ give rise to an overestimation of the unit cell volume, which is a well known trend \cite{goh17,huang12}. For a set of $U_{\mathrm{O}-2p}=U_{\mathrm{Zn}-3d}=8$~eV used for subsequent electronic structure calculations the corresponding error is of the order of one tenth of a percent. Results in figure~\ref{fig1new} also suggest that larger values would give better lattice parameters, but we notice that these large values lead to worse results, when it comes to the electronic structure and bandgap.

\subsection{Numerical results}

\begin{figure}[h!]
\centering
\includegraphics{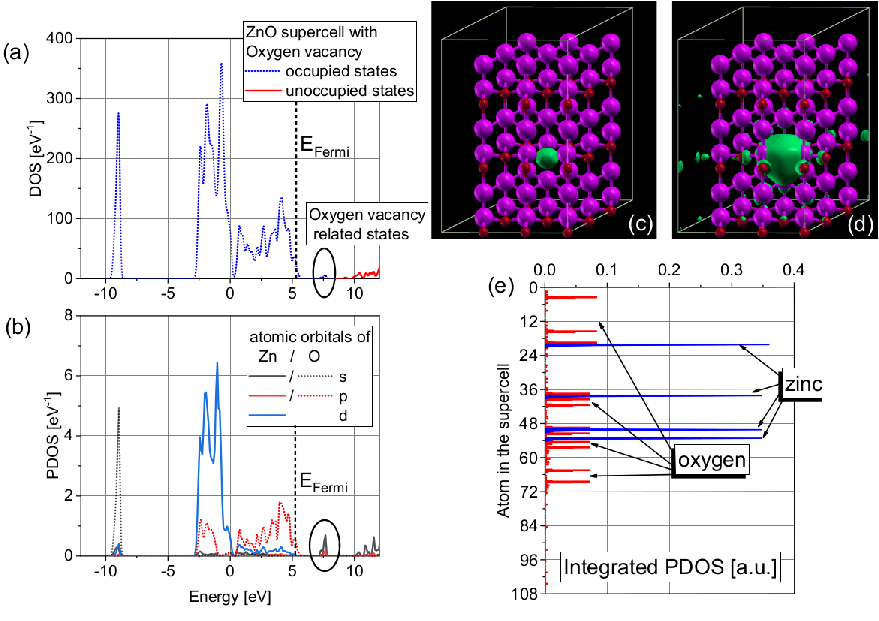}
\caption{Total density of states (DOS) (a) and atomic orbital projected density of states (PDOS) (b) for a $3\times3\times3$ supercell with an oxygen vacancy center (V$_\mathrm{O}$). The Fermi level of defect-free ZnO is shown by a dashed vertical line. Occupied and unoccupied molecular orbitals of DOS are marked by blue dotted and red solid line in (a) correspondingly, whereas the black ellipse indicates the position of the vacancy center. PDOS on $s$ (black), $p$ (red), and $d$ (blue) atomic orbitals of O and Zn --- which belong to the first coordination sphere of V$_\mathrm{O}$ --- are highlighted by dotted and solid lines in (b) respectively. Electronic density of the defect level at $\Gamma$-point in a ZnO supercell with 107 atoms and sampled at 35\% (c) and 3\% (d) of the maximum value is shown by green isosurfaces. Red and magenta spheres in (c)-(d) stand in the positions of oxygen and zinc accordingly. One can notice that the defect state is highly localized at V$_\mathrm{O}$. While the dominant contribution to the electronic density of V$_\mathrm{O}$ in the $3\times3\times3$ supercell comes from four closest to the vacancy center Zn atoms --- blue peaks in (e) --- and their twelve oxygen satellites, i.e. three O atoms per each Zn, --- red peaks in (e).}
\label{fig2new}
\end{figure}

The density of states (DOS) evaluated for a $3\times3\times3$ supercell of ZnO in the presence of an oxygen vacancy, within the pseudohybrid Hubbard DFT approach with $U_{\mathrm{O}-2p}=U_{\mathrm{Zn}-3d}=8$~eV, is shown in figure~\ref{fig2new}(a). Besides the structure of (un)occupied states, which it inherits from the defect-free ZnO, the DOS in figure~\ref{fig2new}(a) accommodates a localized level at 7.5~eV, which is present only after introducing an oxygen vacancy in the supercell. It turns out that this value is very close to the position reported earlier in Ref.~\cite{paudel08}. Although, in Ref.~\cite{paudel08} Hubbard type correction was included to the Zn $s$ orbital which resulted in a bandgap opening, no direct physical interpretation was provided \cite{ma13}. In the present investigation we report also on the spatial extent and geometry of the electronic density isosurfaces (as implemented with \texttt{XCrySDen} visualization tools \cite{xcrysden03}). These results shown in figures~\ref{fig2new}(c)-(d) clearly demonstrate that the defect level is localized on the vacant oxygen site of the supercell, see figure~\ref{fig2new}(c). Thus, the change of the total charge density induced by the vacancy, that we refer to as pseudocharge density, is distributed over the oxygen vacancy center with minor localization on nearest atoms of Zn and O, see figure~\ref{fig2new}(d). We note that for this figure, we sampled $|\psi_D(\bm{r})|^2=const$, with $\psi_D(\bm{r})$ standing for the defect state, at 35\%, figure~\ref{fig2new}(c), and 3\%, figure~\ref{fig2new}(d), of the maximum value.

\begin{figure}[h]
\centering
\includegraphics[scale=0.4,clip]{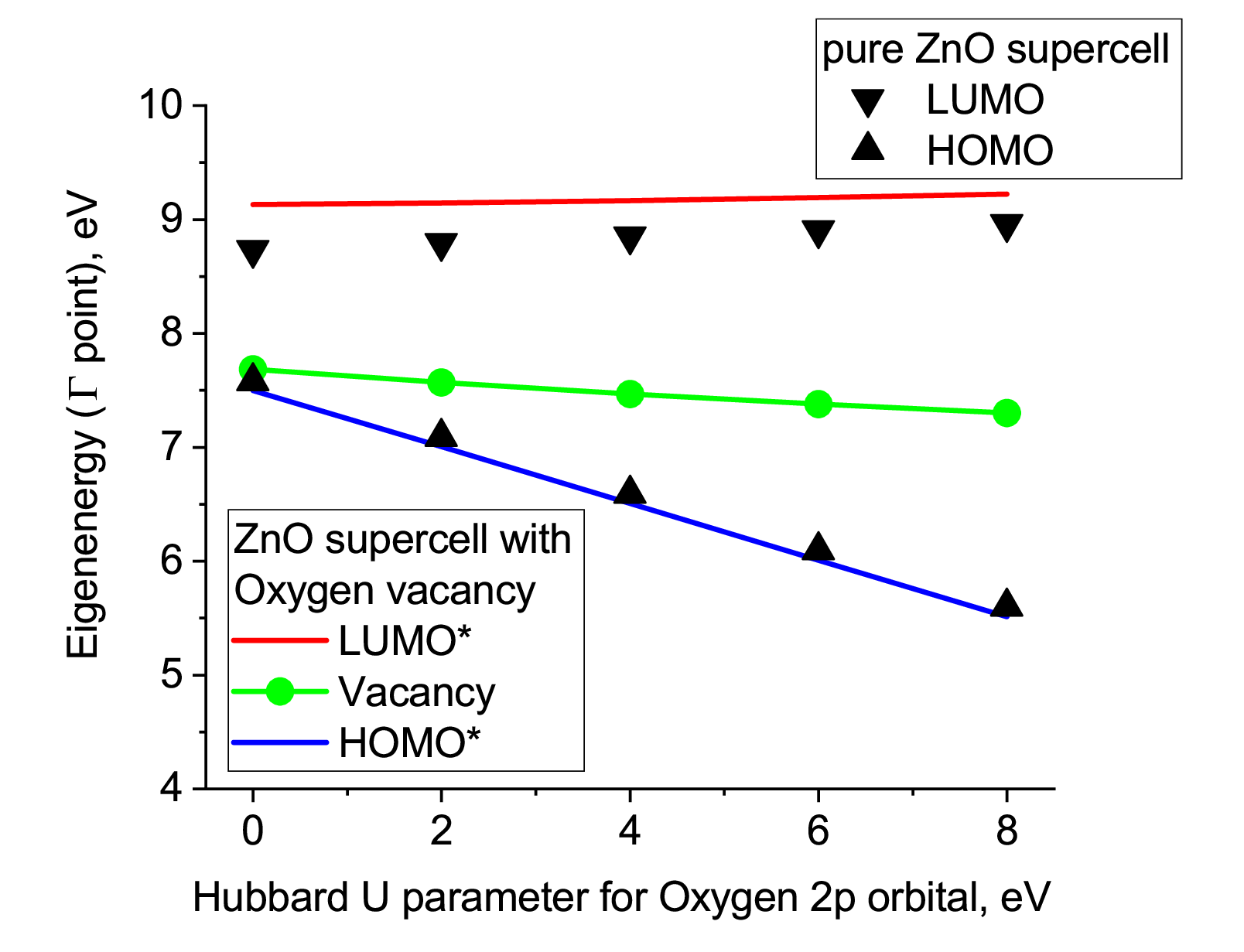}
\caption{The effect of $U_{\mathrm{O}-2p}$ on the ZnO supercell electronic structure with (solid lines) and without (triangled lines) oxygen vacancy for fixed $U_{\mathrm{Zn}-3d}=8$~eV.}
\label{seriesPBE}
\end{figure}

The analysis of the defect level in terms of atomic orbitals can be resolved by inspecting the PDOS, shown in figure~\ref{fig2new}(b). The peak at $-9$~eV corresponds to the $2s^2$ orbital of oxygen, while the states of the zinc $3d^{10}$ shell are located between $-2.5$ and $0$~eV. Remarkably, the $2p$ states of oxygen ($0-6$~eV) are decoupled (primarily by means of the $U_{\mathrm{Zn}-3d}$ Hubbard correction) with those of zinc $3d$ that form the top of the valence band. It should be noted that wave function overlap causes hybridization between the Zn $3d$ and O $2p$ states. As one can see from figures~\ref{fig2new}(a)-(b), the unoccupied orbitals, moving 3.4~eV away from the valence band, are governed by the Zn $4s$ orbitals. The estimated bandgap value reaches 3.5~eV which is close to the experimental result of 3.4~eV. 

To check the stability of the obtained results, we study the dependence of highest occupied (HOMO) and lowest unoccupied (LUMO) molecular orbital position of ZnO with and without oxygen vacancy on $U_{\mathrm{O}-2p}$ parameter shown in figure~\ref{seriesPBE}. It is clearly visible that the bandgap of doped with vacancy ZnO supercell grows while increasing of $U_{\mathrm{O}-2p}$ parameter. In the absence of $U_{\mathrm{O}-2p}$ correction,  the energy of defect state coincides with HOMO level resulting in the bandgap value of 1.3~eV which coherent with previous results \cite{Oba10,janotti06}. It is clear that the vacancy level in ZnO is primarily achieved by the influence of the Hubbard correction applied to the oxygen $2p$ states, $U_{\mathrm{O}-2p}$, and it is robust against variations of the highlighted parameter. 
We also integrate the PDOS from 107 atoms of the $3\times3\times3$ ZnO supercell with an oxygen vacancy center over the energy interval of the defect state [$7-8$~eV in figure~\ref{fig2new}(a)] and sum for all orbitals of each atom. The following allows one to obtain the atomic resolved information that is depicted in figure~\ref{fig2new}(e). Remarkably, figure~\ref{fig2new}(e) shows no qualitative difference with the case of defect-free ZnO, except for the atoms forming the two closest coordination spheres of the oxygen vacancy center (four blue peaks from Zn atoms and twelve red peaks from their O satellites in figure~\ref{fig2new}(e)). The obtained result is consistent with spatial configuration of electronic density of the defect state localized on atoms, which are close to the vacancy center (figures~\ref{fig2new}(c)-(d)). Moreover, it is interesting to note that the defect level overlap with atomic wave functions is about 90\%.

\subsection{State decomposition}

As reported in Ref.~\cite{lany08} the defect state, $\psi_D(\bm{r})$, can be decomposed in the basis of those from the top of the valence band and the bottom of the conduction band. In our study, the defect level for a neutral vacancy, V$^0_\mathrm{O}$, appears due to Hubbard correction, applied to oxygen $2p$ orbitals and is accompanied by a gap opening. In Ref.~\cite{lany08}, however, the gap opening is induced by a Hubbard correction to the Zn $s$ orbital, which in this case quenches the physically relevant $U_{\mathrm{Zn}-3d}$ correction. Moreover, implementation of Zn $s$ correction did not allow to optimize supercell geometry \cite{lany08}. In contrast, as outlined in Ref.~\cite{agapito15}, the physically relevant correction to the oxygen $2p$ states does not corrupt the $d$-orbital state, yields a good equilibrium geometry of the ZnO supercell, and opens up a gap, which is compatible to experimental results.

\begin{figure}[h!]
\centering
\includegraphics{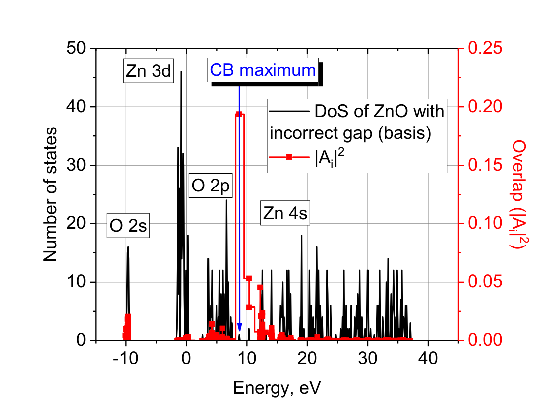}
\caption{Expansion of the localized state, associated with the oxygen vacancy center, in terms of molecular orbitals of the ZnO supercell (with no $U_{\mathrm{O}-2p}$ included) at the $\Gamma$ point of the Brillouin zone. Number of states shown on left axis and overlap (see text) shown on the right axis.}
\label{fig4new}
\end{figure}

Since the PDOS of the defect state is mainly formed by the nearest zinc orbitals the contribution of the valence band states to the defect level is unlikely, because the top of the valence band is formed by oxygen states. Taking this into consideration, we expand the defect states, $\psi_D(\bm{r})$, in terms of all explicitly calculated 486 occupied and 506 unoccupied molecular orbitals of the defect-free sample, $\psi_i(\bm{r})$, in the vicinity of the $\Gamma$-point. This allows to calculate amplitudes of the decomposed states as: $A_i|_{\bm{k}=\Gamma}=\int_{V}d \bm{r}\,\psi_i^\ast(\bm{r})\psi_D (\bm{r})$, integrated over the volume of the supercell, $V$. A close inspection of the results of numerical integration, shown in figure~\ref{fig4new}, reveal that the defect state is mainly constituted by conduction states of the defect-free ZnO supercell. A summation over $\sum_i|A_i|^2\approx0.98$, indicates that the calculated single particle wave functions represent almost complete basis set.

\subsection{Effective mass calculations}

\begin{figure}[h!]
\centering
\includegraphics[scale=0.8]{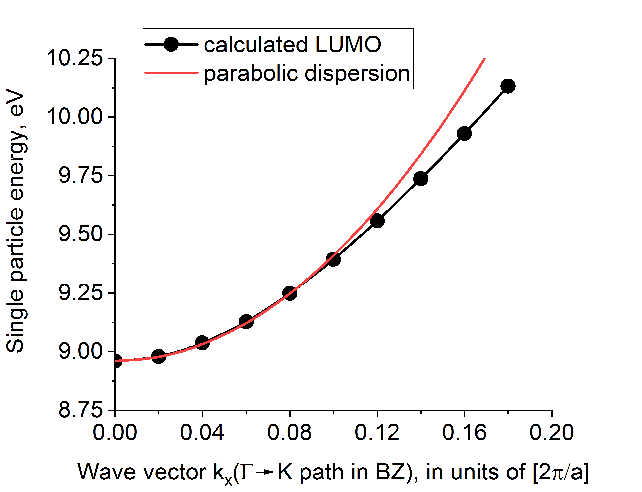}
\caption{Estimation of the effective mass of charge carriers (along $\Gamma\rightarrow K$ direction) by fitting parabolic dependence into band structure of ZnO calculated by means of the fully Hubbard corrected method.}
\label{fig5new}
\end{figure}

Fitting the band structure in the vicinity of high-symmetry $k$ points of the Brillouin zone with a parabolic dispersion allows one to qualitatively evaluate the effective mass of charge carriers \cite{morkoc09}. Typically, switching on a Hubbard correction opens up a bandgap with no significant impact on band curvature \cite{goh17,agapito15, flores18}. The results of our numerical calculations in the vicinity of the $\Gamma$-point are shown in figure~\ref{fig5new}. Note that the resulting dispersion is well approximated by $k^2/m_\ast$ with the effective mass of $0.3\,m_e$ (here $m_e$ is a free electron mass). Interestingly, calculated effective masses along $k_x$ and $k_y$ directions manifest similar values, while in $k_z$ direction the estimation gives slightly (2\%) lower value. In experimentally accessible regime, this value ranges from $m_\ast=0.23\,m_e$ \cite{baer67,imanaka01} to $m_\ast=0.28\,m_e$ \cite{button72} (or even $m_\ast=0.34\,m_e$ \cite{button72}), which is purely determined by experimental procedure and sample preparation method. The results of first-principles calculations within homogeneous electron gas approximation report the effective mass to be $m_\ast\sim0.13-0.14\,m_e$ \cite{karazhanov06}, while GWA corrected band structure resolves $m_\ast=0.24\,m_e$ \cite{oshikiri01}.

\section{Conclusions}

\begin{figure}[h!]
\centering
\includegraphics{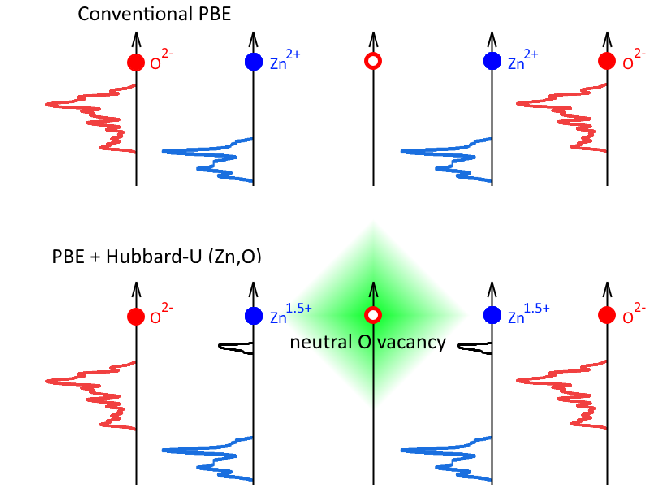}
\caption{Schematic representation of the Hubbard corrections effect on the calculated electronic structure of the oxygen vacancy in a ZnO supercell. The oxygen vacancy center is surrounded by four Zn atoms (two of them are shown), each of which possesses three O satellites. Obviously, a necessity of Zn $3d$ correction can be easily validated, as it allows to align calculated DOS according to the measured XPS spectra representing binding energy of localized $d$ electrons. Whereas the use of the Hubbard correction to O $2p$ orbital is more tricky, however it is shown by outcomes and direct verification \cite{agapito15} that this method is reliable and valuable. Electronic density at the oxygen vacancy center is provided by the $s$ shells of the nearest Zn atoms.}
\label{fig6new}
\end{figure}

In this work, we have provided a comprehensive and self-consistent calculation of an oxygen vacancy in a ZnO supercell, within the framework of a pseudohybrid Hubbard DFT approach. The optimal values of the $U$ parameters in the simplified DFT+$U$ method \cite{cococcioni05,himmetoglu14,dudarev98} for both Zn and O atoms are set in line with recently calculated results \cite{agapito15}; and have been further verified by virtue of a direct comparison of the calculated effective mass of charge carriers and relaxed lattice constants with available experimental data. In contrast to the previous studies based on homogeneous electron gas approximation, the method proposed here can be used for modelling the ground state electronic structure of the neutral oxygen vacancy in ZnO for large supercells. The effective concentration of oxygen vacancies in our simulations is rather high (up to 10$^{-21}$~cm$^{-3}$), thus a small dispersion of the defect level is observed from these calculations. Due to neutrality conditions, the defect state has an effective charge that is equal to -2, where the main contribution to the electronic density comes from the $s$ shell of the nearest Zn atoms (as schematically outlined in figure~\ref{fig6new}). Based on an analysis of wave function decomposition, we observe that the defect state is due to on-site interaction, that removes the excessive attraction to the bulk electronic density of a valence band and results in dissolving the defect state of the vacancy in the valence band. It occurs due to the fact that the electronic structure of the defect is mainly composed of the Zn $4s$ shell and is not related to the valence band states.

As a final set of analysis, we discuss the results of available magneto-resonance measurements \cite{Vlasenko2010} for determining the position of a localized level, associated with an oxygen vacancy. In particular, analyzing photosensitivity of neutral (V$_\mathrm{O}^0$) and singly positively charged (V$_\mathrm{O}^+$) oxygen vacancies by virtue of the photo-electron paramagnetic resonance method, suggests for the V$_\mathrm{O}^0$, an F-center, to be positioned 1.2~eV above the valence band with photoionization energy of 2.3~eV \cite{Nikitenko2001}. While the results of optically-detected magnetic resonance studies place this donor level to 0.9~eV above the valence band edge \cite{Vlasenko2005,Vlasenko2006}.
The results of experimental studies on effective mass determination and defect level position are quite sensitive to tiny details of experiment and strongly depend on quality of the samples, which explains the slight scatter in experimental values. The calculated values obtained here from the pseudohybrid Hubbard DFT approach is 1.5~eV (for the defect level with respect to the top of the valence band), which is in rather good agreement with experimental observations. This implies that the introduced method allows to reproduce the crystal geometry and the electronic structure, including the bandgap of defect-free ZnO, as well as the position of defect levels associated with oxygen vacancies. 

\ack
I.I.V. acknowledges the support from Russian Foundation for Basic Research Project No. 20-52-S52001. A.A.P. acknowledges the support from Russian Foundation for Basic Research Project No. 19-32-60020. The work of D.Y. was supported by the Swedish Research Council (Vetenskapsr{\aa}det, 2018-04383). P.A.R. acknowledges the support from the Russian Foundation for Basic Research Project 18-52-76002. O.E. also acknowledges support from the Knut and Alice Wallenberg foundation, eSSENCE and STandUPP. The work of O.E. and B.S. was supported by the Swedish Research Council.

\section*{References}
\bibliographystyle{iopart-num}
\bibliography{main.bbl}

\end{document}